\documentclass[a4paper,12pt]{article}

\usepackage{geometry}
\usepackage{amsmath}
\usepackage{amssymb}
\usepackage{graphicx} 
\usepackage{epstopdf} 
\usepackage[font=footnotesize,skip=10pt,justification=centering,belowskip=-10pt,aboveskip=10pt]{caption}
\usepackage{subcaption}
\usepackage{multirow} 
\usepackage{textcomp} 
\usepackage{stmaryrd}
\usepackage{makeidx}
\usepackage{epsfig}
\usepackage{float}
\usepackage{wasysym} 
\usepackage{authblk} 
\usepackage{algorithm2e}

\DeclareMathOperator\arctanh{arctanh}

\title{\bf{Efficiency and accuracy of GPU-parallelized Fourier spectral methods for solving phase-field models}}

\author[a,b,c,*]{A.D.~Boccardo}
\author[a,b]{M.~Tong}
\author[a,b]{S.B.~Leen}
\author[c]{D.~Tourret}
\author[c,d,*]{J.~Segurado}

\affil[a]{\small{Mechanical Engineering, School of
    Engineering, College of Science and Engineering, University of Galway,
    University Road, Galway H91 HX31, Ireland.}}

\affil[b]{\small{I-Form Advanced Manufacturing Research Centre,
    University of Galway, University Road, Galway H91 HX31, Ireland.}}

\affil[c]{\small{IMDEA Materials Institute, C/ Eric Kandel 2,
    28906, Getafe, Madrid, Spain. $^{*}$email
    corresponding author: adrian.boccardo@imdea.org}}

\affil[d]{\small{Universidad Polit\'ecnica de Madrid, Department
    of Materials Science, E.T.S.I. Caminos, C/ Profesor Aranguren 3,
    28040, Madrid, Spain. $^{*}$email corresponding author: javier.segurado@imdea.org}}

\date{ }

\begin{document}
\maketitle

\begin{abstract}
  Phase-field models are widely employed to simulate microstructure evolution during processes such as solidification or heat treatment. 
  The resulting partial differential equations, often strongly coupled together, may be solved by a broad range of numerical methods, but
  this often results in a high computational cost, which calls for advanced numerical methods to accelerate their resolution. Here, we
  quantitatively test the efficiency and accuracy of semi-implicit Fourier spectral-based methods, implemented in Python programming language
  and parallelized on a graphics processing unit (GPU), for solving a phase-field model coupling Cahn-Hilliard and Allen-Cahn equations. We compare computational performance and accuracy with a standard explicit finite difference (FD) implementation
  with similar GPU parallelization on the same hardware. For a similar spatial discretization, the semi-implicit Fourier spectral (FS) solvers
  outperform the FD resolution as soon as the time step can be taken 5 to 6 times higher than afforded for the stability of the FD scheme. The
  accuracy of the FS methods also remains excellent even for coarse grids, while that of FD deteriorates significantly. Therefore, for an equivalent
  level of accuracy, semi-implicit FS methods severely outperform explicit FD, by up to 4 orders of magnitude, as they allow
  much coarser spatial and temporal discretization.
\end{abstract}

\vspace{1pc}

%
\section{Introduction}
%
The Phase-Field (PF) method is extensively employed to simulate microstructure evolution during solidification and solid-state
transformations of alloys, as well as many other problems involving complex pattern formation and their
evolution \cite{chen2002phase, boettinger2002phase, moelans2008introduction, steinbach2009phase, tonks2019phase, TOURRET2022100810}. 
PF models consist of one or several coupled partial differential equations that are solved, in general, in 2D or 3D domains. 
The resolution may be implemented with a range of
numerical methods, e.g., finite difference (FD), finite element, finite volume, and Fourier spectral (FS), and different hardware, such as central processing units (CPUs) and graphics processing units (GPUs). However, the need for a fine resolution at the scale of
the smallest morphological details, e.g., at the scale of dendrite tips for solidification, often results in high computational cost,
particularly when large 3D domains are considered. During the last couple of decades, numerous strategies have been explored to
accelerate PF calculations. 

The most widely used numerical method to solve PF models considers a standard first-order forward Euler finite difference (explicit)
for time discretization and second-order finite difference for space discretization. Performance improvement may be achieved by
means of parallel computing. Notably, George and Warren \cite{george2002parallel}, as well as Nestler and colleagues
\cite{NESTLER2005e273,Vondrous_2014} employed a multi-CPU parallelization by means of message passing interface (MPI). Provatas
\textit{et al.} \cite{PROVATAS1999265,GREENWOOD2018153} used an adaptive mesh refinement algorithm together with multi-CPU
parallelization. Non-uniform meshing allows locally refined grid size at the interface and, in this way,  it is possible to
reduce the total number of degrees of freedom. The extra time required for the remeshing process is usually compensated for
problems in which the (refined) interface region represents a small subset of the total domain. Mullis, Jimack and colleagues
\cite{rosam2008adaptive} employed a second-order fully implicit scheme, along with adaptive mesh refinement, and multi-CPU
\cite{BOLLADA2015130}. Takaki and colleagues \cite{YAMANAKA201140,Sakane_2015} combined the compute unified device architecture
(CUDA) and massive parallelization over hundreds of GPUs and CPU cores. Advanced multi-GPU strategies \cite{Shimokawabe_2011},
most recently combined with adaptive mesh refinement and load balancing algorithms \cite{sakane2022parallel}, have led to PF
simulations among the largest reported to date. Such implementations have enabled to tackle challenging multiscale
problems, for instance applied to three-dimensional simulations of solidification considering both thermal and solute fields
\cite{BOLLADA2015130}, polycrystalline cellular/dendritic growth \cite{takaki2018competitive, song2023cell}
or eutectic growth \cite{hotzer2015large} at experimentally-relevant length and time scales.

A very common alternative to the finite element method or FD in many types of microscopic simulations is the use of Fourier
spectral solvers due to their intrinsic periodic nature, which is often an acceptable assumption for microscopic problems at the scale of representative volume elements, and due to their
superior numerical performance \cite{Lucarini2021}. The reason behind the efficiency of FS solvers is the use of the fast Fourier transform (FFT)
algorithm to perform the Fourier transforms at a computational cost that scales with complexity as $N\log(N)$, where $N$ is the amount of data input. 
Regarding their
application to PF, semi-implicit Fourier spectral methods (often named directly FFT methods) have been proposed to solve PF
models with periodic boundary conditions (BCs) \cite{CHEN1998147, zhu1999coarsening, FENG2006498}.  In contrast to popular Euler FD schemes, their
semi-implicit nature allows the use of larger time steps in comparison with explicit schemes. Moreover, the spatial discretization
in the Fourier space gives an exponential convergence rate, in contrast to second order offered by the above-mentioned finite
difference method.  
Moreover, FFTs may be used to solve general FD schemes in Fourier space, while keeping the semi-implicit nature of the integration. 
As a result, FS resolution of
phase-field models were found to offer excellent accuracy and acceleration by orders of magnitude compared to the usual explicit
FD algorithms, with particular efficiency for problems involving long-range interactions \cite{CHEN1998147, zhu1999coarsening}.
To further improve computational efficiency, semi-implicit FS method have also been combined with adaptive meshing, using a
non-uniform grid for the physical domain, a uniform mesh for the computational one, and a time-dependent mapping between
them \cite{FENG2006498}. 

Thanks to their outstanding performance, semi-implicit FFT-based implementations have been used beyond standard PF simulations.
For example, they have a very good potential to solve problems coupling microstructural evolution with micromechanics
\cite{chen2015integrated} taking advantage of the similar frameworks used for phase-field and crystal plasticity
models. Another example, is the use of FFT solvers for phase-field crystal (PFC) models \cite{cheng2008efficient, tegze2009advanced}.

Independently of the intrinsic performance of the numerical method, massive parallelization is a fundamental ingredient
for efficient PF simulations of large domains. As previously stated, CPU and GPU approaches have been widely
explored in FD implementations of PF models. Regarding spectral solvers, first implementations were developed for CPU
parallelization \cite{CHEN1998147, zhu1999coarsening} while most recent developments rely on GPU parallelization showing
great potential \cite{Shi_2017,LEE2019109088,eghtesad2018graphics}.  However, rigorous analyses of the efficiency and accuracy
of combined Fourier based methods and GPU parallelization on reference benchmark problems are lacking. 

In this paper, we perform a quantitative assessment of the accuracy and efficiency of a semi-implicit Fourier
spectral method parallelized on GPU compared to the most common standard approach in PF, i.e., an explicit FD scheme with similar GPU
parallelization. This comparison is made for a typical phase-field benchmark problem representative of the Ostwald ripening
phenomenon. The Fourier approaches include both a \emph{pure} spectral approach and several different FD schemes,
all integrated in time using a semi-implicit scheme. 
First, we summarize the phase-field problem formulation in Section \ref{spf}.
The semi-implicit Fourier spectral resolution scheme is presented in Section \ref{sfftbr}. In Section \ref{srd}, the resulting
accuracy and computational performance are compared with that of a standard explicit finite difference
scheme, also implemented in Python and parallelized on a single GPU. Finally, we summarize our conclusions in Section \ref{sc}.

%
\section{Methods}
%

\subsection{Problem formulation}
\label{spf}
We use the Ostwald ripening phase-field simulation benchmark proposed in Ref.~\cite{jokisaari_2017} to test our FS-GPU
implementation. The simulation represents the growth and coarsening of $p=4$ variants of $\beta$ particles into an $\alpha$ matrix. The
total free energy of the system formed by $\alpha$ and $\beta$ phases
is \cite{jokisaari_2017, wheeler1992phase, zhu2004three}:

\begin{equation}
  F=\int_{V} \Big{(} f_{chem}+\dfrac{\kappa_{c}}{2} |\nabla c|^{2} + \sum_{i=1}^{p} \dfrac{\kappa_{\eta}}{2} |\nabla \eta_{i}|^{2} \Big{)} dV
  \label{eq:fe}
\end{equation}
where $f_{chem}$ is the chemical free energy density, $c$ is the composition field,
$\eta_{i}$ are order parameters (i.e., the phase fields), $\kappa_{c}$ and $\kappa_{\eta}$ are the gradient
energy coefficients for $c$ and $\eta$, respectively, and $V$ is the volume. 
Regions in the domain where \{$\eta_i=0$, $\forall \ i$\} correspond to the $\alpha$ matrix, while regions where \{$\eta_{i}=1$ and
$\eta_{j}=0$, $\forall \ j\neq i$\} correspond to $\beta$ particles of variant $i$ (e.g., grain orientation index).

The chemical free energy density is defined as $f_{chem}=f^{\alpha}(1-h)+f^{\beta}h+w g$,
where $f^{\alpha}=\varrho^{2}(c-c_{\alpha})^{2}$ and $f^{\beta}=\varrho^{2}(c_{\beta}-c)^{2}$
are the chemical free energy densities of $\alpha$ and $\beta$ phases, respectively,
$\varrho$ parametrizes the concentration-dependence of the free energies,
$c_{\alpha}$ and $c_{\beta}$ are the equilibrium concentrations of $\alpha$ and $\beta$
phases, respectively, $h$ is an interpolation function, $g$ is a double-well function,
and $w$ is a parameter that controls the height of the double-well barrier.
The interpolation and double-well functions are defined as
$h=\sum_{i=1}^{p} \eta_{i}^{3}(6\eta_{i}^{2}-15\eta_{i}+10)$ and
$g=\sum_{i=1}^{p}[\eta_{i}^{2}(1-\eta_{i})^{2}]+\alpha \sum_{i=1}^{p} \sum_{j \ne i}^{p}(\eta_{i}^{2}\eta_{j}^{2})$,
where $\alpha$ is a parameter that prevents the overlapping of different non-zero $\eta_{i}$.

The evolution of the $c$ and $\eta_{i}$ fields are computed with Cahn-Hilliard \cite{cahn1961spinodal} and Allen-Cahn
\cite{allen1979microscopic} partial differential equations, respectively:

\begin{align}
  \dfrac{\partial c}{\partial t} &= \nabla \cdot \Big{\{} M \nabla \Big{(} \dfrac{\partial f_{chem}}{\partial c} -\kappa_{c}\nabla^{2}c\Big{)}\Big{\}}
  \label{eq:C-H}
\\
  \dfrac{\partial \eta_{i}}{\partial t} &= -L\Big{(}\dfrac{\partial f_{chem}}{\partial \eta_{i}} -\kappa_{\eta}\nabla^{2}\eta_{i}\Big{)}
  \label{eq:A-C}
\end{align}
where $M$ is the mobility of the solute and $L$ is a kinetic coefficient.

Eqs. \eqref{eq:C-H} and \eqref{eq:A-C} are solved by imposing periodic BCs. 
Following the two-dimensional (2D) benchmark for square domains of size 200 presented in \cite{jokisaari_2017}, $\kappa_{c}=3$,
$\kappa_{\eta}=3$, $\varrho^{2}=2$, $c_{\alpha}=0.3$, $c_{\beta}=0.7$, $w=1$, $\alpha=5$,
$M=5$, $L=5$, and $p=4$. The concentration and order parameter fields are initialized with:

\begin{align}
  c(x,y) = c_0 + \epsilon \, &\big\{\cos(0.105x)\cos(0.11y)+\left[\cos(0.13x)\cos(0.087y)\right]^2 \nonumber\\
  &+ \cos(0.025x-0.15y)\cos(0.07x-0.02y) \big\}
  \label{eq:ic_c}
  \\
  \eta_i(x,y) = \eta_\nu \, &\big\{\cos((0.01i)x-4)\cos((0.007+0.01i)y) \nonumber\\
  & + \cos((0.11+0.01i)x)\cos((0.11+0.01i)y) \nonumber\\
  & + \psi \big[ \cos((0.046+0.001i)x \nonumber\\
  & + (0.0405+0.001i)y)\cos((0.031+0.001i)x \nonumber\\
  & - (0.004+0.001i)y) \big]^2 \big\}^2
  \label{eq:ic_eta}
\end{align}
where $c_{0}=0.5$, $\epsilon=0.05$, $\eta_{\nu}=0.1$ and $\psi=1.5$.
These initial conditions were chosen to lead to nontrivial solutions \cite{jokisaari_2017}.
However they lack the spatial periodicity required to rigorously test and compare them using
periodic BCs, which is also essential to apply the FFT-based solver.
In order to address the lack of periodicity, we let the system evolve, applying Eqs~\eqref{eq:ic_c} and \eqref{eq:ic_eta} as initial conditions and periodic BCs,
for a dimensionless time from 0 to 1, using finite differences with a fine spatial grid and small
time step. The resulting periodic fields at $t=1$ are used as initial condition for all the cases studied.

In addition to a thorough investigation of the 2D benchmark as proposed in \cite{jokisaari_2017}, we also explore a three-dimensional (3D) test case extrapolated from the original case.
In this case, we consider cubic domain of edge size 100.
The phase fields $\eta_i$ are initialized to the same 2D field at $t=1$ as in 2D cases along the $(z=l_z/2)$ plane, with $l_z=100$ the height of the domain, and they are extrapolated along the $z$ direction to produce 3D shapes using the following function:
\begin{align}
  \eta^{(3\rm D)}_{i}(x,y,z) &= 0.5 \bigg{\{}1+\tanh \bigg{[}\arctanh \bigg{(} 2\eta^{(2\rm D)}_{i}(2x,2y)-1 \bigg{)} -\bigg{(}\dfrac{z-l_{z}/2}{5}\bigg{)}^{4} \bigg{]} \bigg{\}}
  \label{eq:ic_eta_3d}
\end{align}
where factors 2 within the $\eta^{(2\rm D)}$ function are due to the rescaling of the domain from 200 to 100 in size. 
The exponent 4 and denominator 5 of the last term of Eq.~\eqref{eq:ic_eta_3d} were simply adjusted to produce $(\eta_i=0.5)$ shapes relatively close to ellipsoids without changing the location of the $(\eta_i=0.5)$ interface in the $(z=l_z/2)$ plane.
The initial condition for the concentration field is set as $c(x,y,z)=c_{0}$ in the entire domain.

%
\subsection{Numerical resolution}
\label{sfftbr}
%
The system of partial differential equations formed by Eqs.~\eqref{eq:C-H} and \eqref{eq:A-C} is solved by means of a
semi-implicit non-iterative Fourier spectral method with first-order finite difference for time discretization. Hence,
${\partial f_{chem}}/{\partial c}$ and ${\partial f_{chem}}/{\partial \eta_{i}}$ are computed considering their value at the
previous time step (explicit part) and the Laplacian of $c$ and $\eta_{i}$ are computed at the current time step (implicit
part). Following the scheme proposed by Chen and Shen \cite{CHEN1998147}, the resolution in the Fourier space, at
time $t+\Delta t$, is:
  
\begin{align}
  c^{(t+\Delta t)}+M \Delta t \kappa_{c} \nabla^{2}\big{(}\nabla^{2}c^{(t+\Delta t)}\big{)} &= c^{(t)}+M \Delta t \nabla^{2}\Bigg{(}\dfrac{\partial f_{chem}^{(t)}}{\partial c} \Bigg{)}
  \label{eq:C-H_discrete}
  \\
  \eta_{i}^{(t+\Delta t)}-L \Delta t \kappa_{\eta} \nabla^{2}\eta_{i}^{(t+\Delta t)} &= \eta_{i}^{(t)}-L \Delta t \dfrac{\partial f_{chem}^{(t)}}{\partial \eta_{i}}
  \label{eq:A-C_discrete}
\end{align}
where $\Delta t$ is the time step. 
For the sake of clarity, the unknown fields $c^{t+\Delta t}$ and $\eta_{i}^{t+\Delta t}$ will
be referred to as $c$ and $\eta_{i}$, respectively.

By definition of the Fourier transform, the gradient of a field $f$ is:

\begin{align}
  \widehat{\nabla f} &= i\boldsymbol{\xi} \widehat{f}
  \label{eq:grad}
\end{align}
with $i$ the imaginary unit, $\boldsymbol{\xi}$ the frequency vector, and $\widehat{~}$ denotes the Fourier transform of the affected variable.
The previous differential equations can be transformed to the Fourier space, resulting in:

\begin{align}
\big{(}1+M \Delta t \kappa_{c} \|\boldsymbol{\xi}\|^4 \big{)} \widehat{c}( \boldsymbol{\xi}) &= \widehat{c}^{(t)}( \boldsymbol{\xi}) -M \Delta t \|\boldsymbol{\xi}\|^2 \widehat{\dfrac{\partial f_{chem}^{(t)}}{\partial c}}( \boldsymbol{\xi}) 
  \label{eq:C-H_discrete_fft}
  \\
\big{(} 1+L \Delta t \kappa_{\eta} \|\boldsymbol{\xi}\|^2 \big{)} \widehat{\eta_{i}} ( \boldsymbol{\xi}) &= \widehat{\eta_{i}}^{(t)}( \boldsymbol{\xi}) -L \Delta t \widehat{\dfrac{\partial f_{chem}^{(t)}}{\partial \eta_{i}}}( \boldsymbol{\xi}) 
  \label{eq:A-C_discrete_fft}
\end{align}
where the expressions in  Eqs.~\eqref{eq:C-H_discrete_fft} and \eqref{eq:A-C_discrete_fft} are two linear systems of algebraical
equations in which the right hand side is the independent term $\mathbf{b}$, depending on the values of the fields
at the previous time step:

\begin{align}
  \widehat{b}_{c}^{(t)}(\boldsymbol{\xi}) &= \widehat{c}^{(t)}-M \Delta t \|\boldsymbol{\xi}\|^2 \widehat{\dfrac{\partial f_{chem}^{(t)}}{\partial c}}
  \label{eq:C-H_fft_b}
  \\
  \widehat{b}_{\eta_{i}}^{(t)} (\boldsymbol{\xi}) &= \widehat{\eta_{i}}^{(t)}-L \Delta t \widehat{\dfrac{\partial f_{chem}^{(t)}}{\partial \eta_{i}}}
  \label{eq:A-C_fft_b}
\end{align}

The equations are decoupled for each frequency, so the system can be solved in Fourier space by inverting the left hand side as: 
\begin{align}
  \widehat{c}( \boldsymbol{\xi}) &=  \big{(}1+M \Delta t \kappa_{c} \|\boldsymbol{\xi}\|^4 \big{)} ^{-1} \widehat{b}_{c}^{(t)} (\boldsymbol{\xi})
  \label{eq:A-C_discrete_fft_c}
  \\
  \widehat{\eta_{i} } ( \boldsymbol{\xi}) &= \big{(} 1+L \Delta t \kappa_{\eta} \|\boldsymbol{\xi}\|^2 \big{)} ^{-1} \widehat{b}_{\eta_{i}}^{(t)} (\boldsymbol{\xi})
  \label{eq:A-C_discrete_fft_eta}
\end{align}
and the fields are readily obtained as the inverse Fourier transform of $\widehat{c}$ and $\widehat{\eta_{i}}$.

If the domain under consideration is rectangular of size $l_x\times l_y$ and it is discretized into a grid containing
$p_x$ and $p_y$ points in each direction, the discrete Fourier frequency vector corresponds to
$\boldsymbol{\xi}=2 \pi [ n_{x}/l_{x},n_{y}/l_{y}]$ and the square of the frequency gradient is:

\begin{align}
  \|\boldsymbol{\xi}\|^2 &= 4 \pi^{2} [(n_{x}/l_{x})^{2}+(n_{y}/l_{y})^{2}]  
  \label{eq:Gamma_fft}
\end{align}
with $n_{x}$ and $n_{y}$ the two-dimensional meshgrid matrices generated with two vectors of the form
$[0, ...,(p_{x}/2),-(p_{x}/2-1),...,-1]$ and $[0, ...,(p_{y}/2),-(p_{y}/2-1),...,-1]$.

An alternative to the standard Fourier spectral approach, first proposed in  \cite{Muller1996} to reduce the aliasing
effect in the presence of non-smooth functions, consists in replacing the definition of the derivative in Eq.~\eqref{eq:grad}
with a finite difference derivative, but computing it through the use of Fourier transform. To this end, the finite difference
stencil under consideration (e.g. backward, forward or central differences) is obtained by transporting the function in
Fourier space using the shift theorem \cite{Muller1996}. This method allows calculation of the spatial derivatives by
considering the local values of the fields, reducing possible oscillations in their values but in general losing accuracy.

This approach results in preserving Eq.~\eqref{eq:grad} for computing the gradient but redefining the frequencies. 
The square of the frequency gradient in Eqs.~\eqref{eq:A-C_discrete_fft_c} and \eqref{eq:A-C_discrete_fft_eta}
is thus redefined as:
\begin{align}
  \|\boldsymbol{\xi}_{c2}\|^2 =& -2 \bigg{\{}\dfrac{\cos(\Delta x \xi_{x})-1}{\Delta x^{2}}+\dfrac{\cos(\Delta y \xi_{y})-1}{\Delta y^{2}}\bigg{\}}
  \label{eq:Gamma_fft-fdO2}
  \\
  \|\boldsymbol{\xi}_{c4}\|^2 =& -\dfrac{1}{6} \bigg{\{}\dfrac{16\cos(\Delta x \xi_{x})-\cos(2\Delta x \xi_{x})-15}{\Delta x^{2}} \nonumber \\
  &+\dfrac{16\cos(\Delta y \xi_{y})-\cos(2\Delta y \xi_{y})-15}{\Delta y^{2}}\bigg{\}}
  \label{eq:Gamma_fft-fdO4}
\end{align}
for centered $\mathcal{O}(\Delta l^{2})$ FD and centered $\mathcal{O}(\Delta l^{4})$ FD, respectively, where $\Delta x$
and $\Delta y$ are the distance between two consecutive grid points in the $x$ and $y$ directions, respectively.
$\Delta l$ indicates the approximation of the spatial discretization in the $x$ ($\Delta l = \Delta x$)
and $y$ ($\Delta l=\Delta y$) directions.

If the domain under consideration is a rectangular parallelepiped of size $l_x \times l_y \times l_{z}$ and it is
discretized into a grid containing $p_x$, $p_y$, and $p_z$ points in each direction, the discrete Fourier frequency
vector corresponds to $\boldsymbol{\xi}=2 \pi [ n_{x}/l_{x},n_{y}/l_{y},n_{z}/l_{z}]$ and the square of the frequency
gradient is:

\begin{align}
  \|\boldsymbol{\xi}\|^2 &= 4 \pi^{2} [(n_{x}/l_{x})^{2}+(n_{y}/l_{y})^{2}+(n_{z}/l_{z})^{2}] 
  \label{eq:Gamma_fft_3d}
\end{align}
with $n_{x}$, $n_{y}$, and $n_{z}$ the three-dimensional meshgrid matrices generated with three vectors of the
form $[0, ...,(p_{x}/2),-(p_{x}/2-1),...,-1]$, $[0, ...,(p_{y}/2),-(p_{y}/2-1),...,-1]$, and
$[0, ...,(p_{z}/2),-(p_{z}/2-1),...,-1]$.

For the case of $\mathcal{O} (\Delta l^{2})$ and $\mathcal{O} (\Delta l^{4})$ FS-FD methods, the square of the
frequency gradient is redefined as:

\begin{align}
  \|\boldsymbol{\xi}_{c2}\|^2 =& -2 \bigg{\{}\dfrac{\cos(\Delta x \xi_{x})-1}{\Delta x^{2}}+\dfrac{\cos(\Delta y \xi_{y})-1}{\Delta y^{2}}+\dfrac{\cos(\Delta z \xi_{z})-1}{\Delta z^{2}}\bigg{\}}
  \label{eq:Gamma_fft-fdO2_3d}
  \\
  \|\boldsymbol{\xi}_{c4}\|^2 =& -\dfrac{1}{6} \bigg{\{}\dfrac{16\cos(\Delta x \xi_{x})-\cos(2\Delta x \xi_{x})-15}{\Delta x^{2}}+\dfrac{16\cos(\Delta y \xi_{y})-\cos(2\Delta y \xi_{y})-15}{\Delta y^{2}} \nonumber\\
  &+\dfrac{16\cos(\Delta z \xi_{z})-\cos(2\Delta z \xi_{z})-15}{\Delta z^{2}}  \bigg{\}}
  \label{eq:Gamma_fft-fdO4_3d}
\end{align}
where $\Delta z$ is the distance between two consecutive grid points along the $z$ direction.

The resolution scheme is presented in Algorithm~\ref{alg:fft-based_solution} for 3D domains. The
computational scheme is implemented using the Python programming language. For each time
step, the computation of discrete Fourier transform ($\mathcal{F}$) and discrete inverse
Fourier transform ($\mathcal{F}^{-1}$) is performed, on the GPU device using Scikit-CUDA
\cite{sk_cuda_2021}. 
To solve Eqs.~\eqref{eq:A-C_discrete_fft_c} and \eqref{eq:A-C_discrete_fft_eta}
on the GPU device, CUDA kernels are programmed through PyCUDA \cite{kloeckner_pycuda_2012}, where
the arrays are defined in double precision.

\RestyleAlgo{ruled}
\SetKwComment{Comment}{/* }{ */}
\begin{algorithm}[hbt!]
  \caption{FS solution algorithm for 3D domain}
  \textbf{Variable initializations:} $\kappa_{c}$, $\kappa_{\eta}$, $\varrho$, $c_{\alpha}$,
    $c_{\beta}$, $w$, $\alpha$, $M$, $L$, $p$, $c_{0}$, $\epsilon$, $\eta_{\nu}$, $\psi$, $lx$, $ly$, $lz$,
    $\Delta x$, $\Delta y$, $\Delta z$, $\Delta t$\;
    Time discretization $[0,r \Delta t,...,t_{max}]$, for $r \in \mathbb{Z}$\;
    Space discretization $x=[0,s \Delta x,...,l_{x}]$, $y=[0,s \Delta y,...,l_{y}]$, $z=[0,s \Delta z,...,l_{z}]$, for $s \in \mathbb{Z}$\;
    $\|\boldsymbol{\xi}\|^2$ with Eqs.~\eqref{eq:Gamma_fft_3d}, \eqref{eq:Gamma_fft-fdO2_3d}, or \eqref{eq:Gamma_fft-fdO4_3d}\;
    \While{$t < t_{max}$}{
      $\widehat{\dfrac{\partial f_{chem}^{(t)}}{\partial c}} \xleftarrow{\mathcal{F}} \dfrac{\partial f_{chem}^{(t)}}{\partial c}$\;
      $\widehat{\dfrac{\partial f_{chem}^{(t)}}{\partial \eta_{i}}} \xleftarrow{\mathcal{F}} \dfrac{\partial f_{chem}^{(t)}}{\partial \eta_{i}}$\;
      $\widehat{b}_{c}^{(t)} = \widehat{c}^{(t)}-M \Delta t \|\boldsymbol{\xi}\|^{2} \widehat{\dfrac{\partial f_{chem}^{(t)}}{\partial c}}$\;
      $\widehat{b}_{\eta_{i}}^{(t)} = \widehat{\eta_{i}}^{(t)}-L \Delta t \widehat{\dfrac{\partial f_{chem}^{(t)}}{\partial \eta_{i}}}$\;
      $\widehat{c}^{(t+\Delta t)} = \big{(}1+M \Delta t \kappa_{c} \|\boldsymbol{\xi}\|^{4} \big{)} ^{-1} \widehat{b}_{c}^{(t)}$\;
      $\widehat{\eta_{i}}^{(t+\Delta t)} = \big{(} 1+L \Delta t \kappa_{\eta} \|\boldsymbol{\xi}\|^{2} \big{)} ^{-1} \widehat{b}_{\eta_{i}}^{(t)}$\;
      $c^{(t+\Delta t)} \xleftarrow{\mathcal{F}^{-1}} \widehat{c}^{(t+\Delta t)}$\;
      ${\eta_{i}}^{(t+\Delta t)} \xleftarrow{\mathcal{F}^{-1}} \widehat{\eta_{i}}^{(t+\Delta t)}$\;  
    }
    \textbf{Output generation}\;
  \label{alg:fft-based_solution}
\end{algorithm}

The GPU-parallelized FS resolution algorithm is compared to a standard explicit FD solver, also parallelized on
GPU through PyCUDA. The FD algorithm uses a common explicit first-order forward Euler time stepping, second-order centered
FD scheme for spatial derivatives, and a 5-point (2D domain) or 7-point (3D domain) stencil Laplacian operator
discretized on a regular grid of square or cubic elements. Within a time iteration, all calculations are performed on the GPU
(device) within three kernels calculating (i) the $\mu=\partial f_{chem}/\partial c -\kappa_{c}\nabla^{2}c$ term from
Eq.~\eqref{eq:C-H}, (ii) $c^{(t+\Delta t)}$ from Eq.~\eqref{eq:C-H}, (iii) $\eta_i^{(t+\Delta t)}$ from Eq.~\eqref{eq:A-C}.
Periodic BCs are applied at the end of each kernel on the calculated field (namely: $\mu$, $c$, and $\eta_i$) using one
extra layer of grid points on each side of the domain. Time stepping is applied at the end of the time loop by
swapping addresses of current $(t)$ and next $(t+\Delta t)$ arrays after the three kernel calls (i.e., on the CPU).
Time-consuming memory copies between GPU (device) and CPU (host) are performed only when an output file of the fields
is required.

Performance comparisons were performed without file output, hence not wasting
any time on file writing. The GPU block size was set to $16\times 16$ for 2D cases and $8 \times 8 \times 8$
for 3D cases, which was found to lead to near-optimal performance for all cases. This algorithm represents a nearly
optimal performance for a single-GPU FD implementation -- and hence a fair comparison for FS-based simulations.

Both FS and FD simulations were performed using a single GPU on a computer with the following hardware features:
Intel Xeon Gold 6130 microprocessor, 187~GB RAM, GeForce RTX 2080Ti GPU (4352 Cuda cores and 11~GB RAM), and
software features: CentOS Linux 7.6.1810, Python 3.8, PyCUDA 2021.1, Scikit-CUDA 0.5.3, and CUDA 10.1 (Toolkit 10.1.243).

\subsection{Simulations}

The developed FS-GPU Python code is employed to simulate the diffusion-driven growth of $\beta$ grains, with 4 crystal orientations,
within an $\alpha$ matrix. The nucleation of $\beta$ is generated by the non-uniform fields of the initial condition.
To test the computational performance and accuracy of the results, the time and space are discretized in different ways for 2D cases.
As shown in Table \ref{tab:cases}, 15 cases are considered for each FS method by defining different combinations of $\Delta t$ and
numbers of grid points ($n$) in regular grids. Furthermore, we also use a FD-GPU Python code, considering an explicit centered FD
method in order to compare its results with the proposed FS methods. The 5 cases solved with FD method are also listed in
Table \ref{tab:cases}.

\begin{table}[ht]
  \begin{center}
    \begin{tabular}{ |c|c|c| } 
      \hline
      & \multicolumn{2}{|c|}{$\mathbf{\Delta t}$} \\
      \hline
      \textbf{Number of grid points} &  FD method & FS methods \\
      \hline
      $128^{2}$  & $8.138 \times 10^{-3}$ & $10^{-4}$; $10^{-3}$; $5 \times 10^{-3}$ \\ 
      $256^{2}$  & $5.086 \times 10^{-4}$ & $10^{-4}$; $10^{-3}$; $5 \times 10^{-3}$\\
      $512^{2}$  & $3.560 \times 10^{-5}$ & $10^{-4}$; $10^{-3}$; $5 \times 10^{-3}$\\
      $1024^{2}$ & $2.225 \times 10^{-6}$ & $10^{-4}$; $10^{-3}$; $5 \times 10^{-3}$\\
      $2048^{2}$ & $1.490 \times 10^{-7}$ & $10^{-4}$; $10^{-3}$; $5 \times 10^{-3}$\\   
      \hline
    \end{tabular}
    \caption{Analyzed cases for 2D FS and FD numerical methods.}
    \label{tab:cases}
  \end{center}
\end{table}

For 3D cases, the computational performance is tested by modifying the spatial discretization.
As shown in Table \ref{tab:cases_3d}, 5 cases are considered for each FS and FD methods by defining different of numbers
of grid points in regular grids.

\begin{table}[ht]
  \begin{center}
    \begin{tabular}{ |c|c|c| }
        \hline
        & \multicolumn{2}{|c|}{$\mathbf{\Delta t}$} \\
        \hline
        \textbf{Number of grid points} &  FD method & FS methods \\
        \hline
        $16^{3}$  & $4.06 \times 10^{-2}$ & $5 \times 10^{-3}$\\ 
        $32^{3}$  & $2.54 \times 10^{-2}$ & $5 \times 10^{-3}$\\
        $64^{3}$  & $3.17 \times 10^{-3}$ & $5 \times 10^{-3}$\\
        $128^{3}$ & $7.94 \times 10^{-5}$ & $5 \times 10^{-3}$\\
        $256^{3}$ & $9.93 \times 10^{-6}$ & $5 \times 10^{-3}$\\   
        \hline
    \end{tabular}
    \caption{Analyzed cases for 3D FS and FD numerical methods.}
    \label{tab:cases_3d}
  \end{center}
\end{table}

\section{Results and discussion}
\label{srd}

The maximum values of $\Delta t$ for the FS methods and the FD method were determined by computational
trial-and-error exploration of the algorithm stability. As expected, the semi-implicit FS algorithm is
more stable than explicit FD, particularly when $n$ is high, as it was possible to employ much
higher values of $\Delta t$. Moreover, as in \cite{CHEN1998147}, the unconditional stability of the
semi-implicit scheme was observed when grid size is increased, such that, unlike with FD, it is not necessary
to decrease $\Delta t$ as $n$ increases. 

Below, we first describe the system and free energy evolution for representative simulations with relatively fine
discretization in both FS and FD methods (Section~\ref{sec:resu:fields}), before discussing computational performance
(Section~\ref{sec:resu:perf}) and accuracy (Section~\ref{sec:resu:accu}) when changing $n$ and $\Delta t$.

\subsection{System evolution}
\label{sec:resu:fields}

\begin{figure}[H]
  \begin{center}
    \includegraphics[scale=0.6]{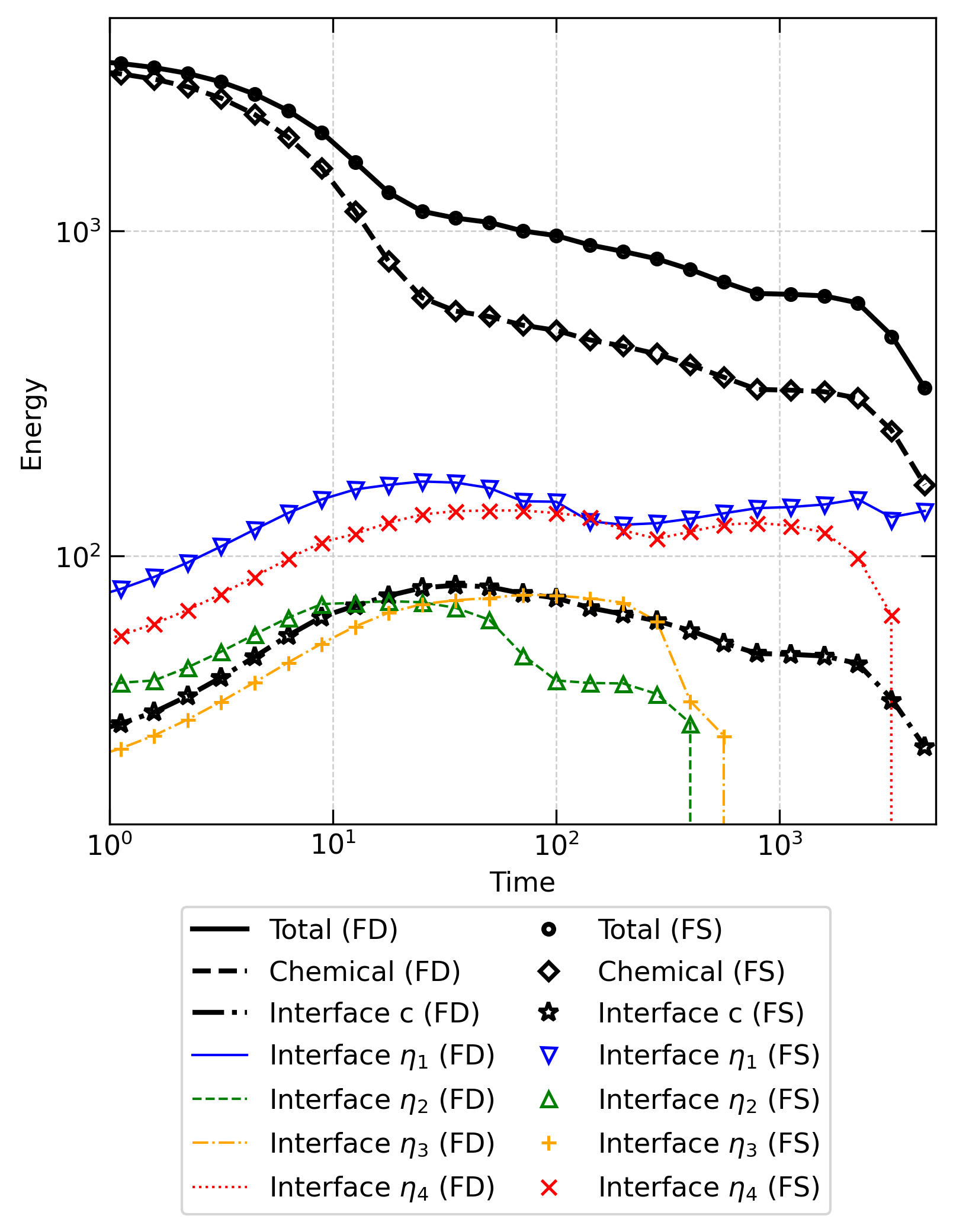}
    \caption{Free energy evolution, for 2D cases, obtained for a $1024^2$ grid with FD ($\Delta t=2.225 \times 10^{-6}$) and
      FS ($\Delta t=10^{-3}$) numerical methods.} 
    \label{fig:free_energy}
  \end{center}
\end{figure}

The evolutions of the total free energy and of its individual components (Eq.~\eqref{eq:fe}), computed with FD
($\Delta t=2.225 \times 10^{-6}$ and $n=1024^{2}$) and FS
($\Delta t=10^{-3}$ and $n=1024^{2}$) methods for 2D cases, are presented in Figure \ref{fig:free_energy}.
A monotonic decrease in total free energy is observed, consistent with the expected energy minimization during
microstructure evolution. 
The chemical contribution (i.e., integrating only the first term in Eq.~\eqref{eq:fe}) behaves similarly as the total free energy because the system tends to
equilibrium. 
The interfacial energy contributions
with respect to $c$ (integrating the second term in Eq.~\eqref{eq:fe})
or $\eta_i$ (integrating independently the last term in Eq.~\eqref{eq:fe} for each $i$)
 increase and decrease when the surface areas of the particles increase and decrease, respectively. 
Fluctuations in $\eta_{i}$ components are attributed to shrinkage of some particles whilst others grow by dissolution
of neighbor particles. 
As expected from the fine grids used here and hence the high accuracy of the two simulations (see Section~\ref{sec:resu:accu}), the two methods predict the
same energy evolution.

\begin{figure}[H]
  \begin{center}
    \includegraphics[scale=0.4]{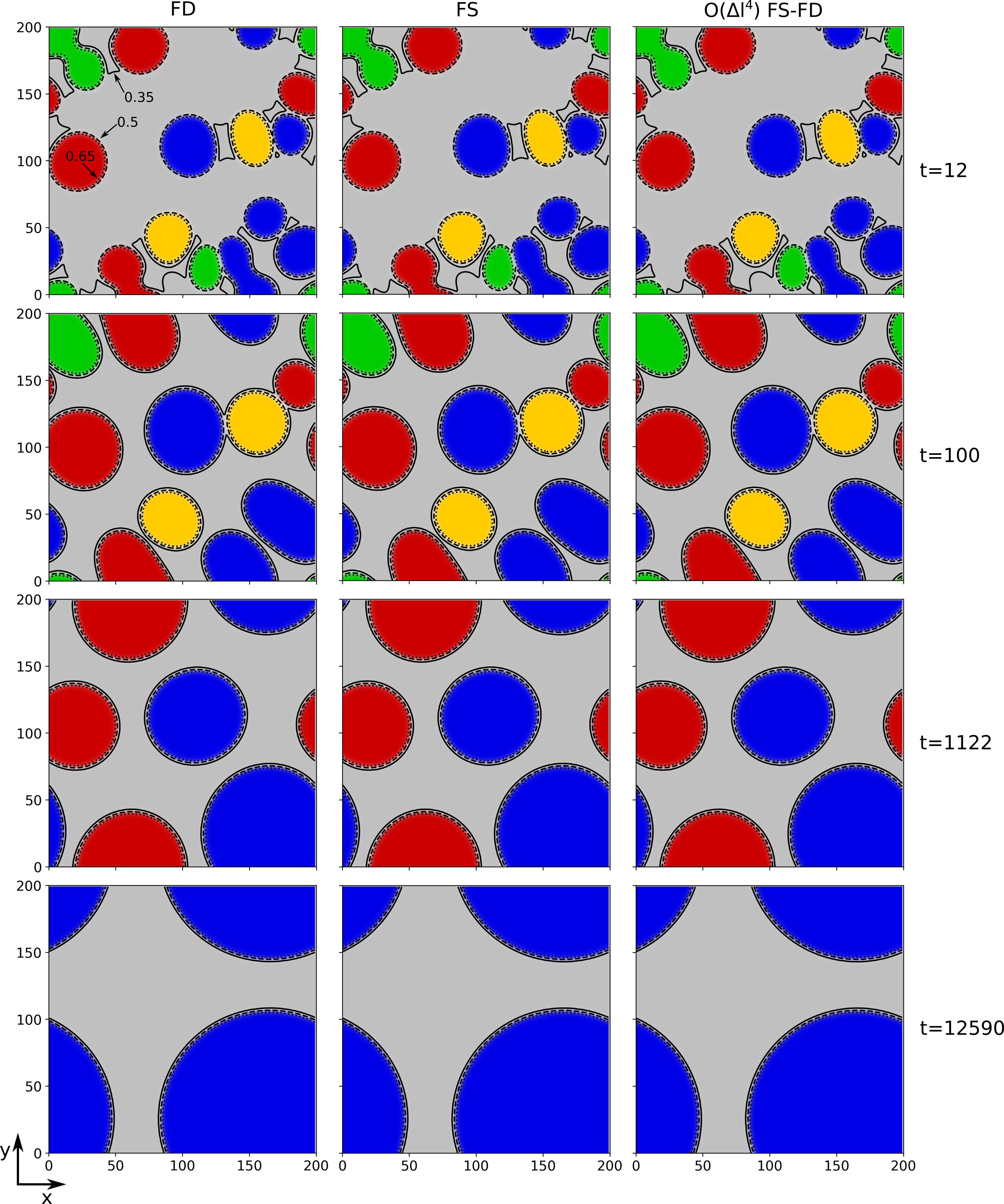}
    \caption{Evolution of phase (color) and concentration (iso-value lines) fields, for 2D cases, at different times (rows)
      for a grid size of $1024^2$ obtained with different methods (columns), namely: (left) FD ($\Delta t=2.225 \times 10^{-6}$);
      (center) FS ($\Delta t=10^{-3}$), and (right) $\mathcal{O} (\Delta l^{4})$ FS-FD ($\Delta t=10^{-3}$).}
    \label{fig:panel_microstructure_2d}
  \end{center}
\end{figure}

Figure \ref{fig:panel_microstructure_2d} shows the evolution of the 2D microstructure simulated on a $1024^2$ grid using FD
($\Delta t=2.225 \times 10^{-6}$), FS ($\Delta t=10^{-3}$), and $\mathcal{O} (\Delta l^{4})$ FS-FD ($\Delta t=10^{-3}$) methods.
The $\eta_{1}$, $\eta_{2}$, $\eta_{3}$, and $\eta_{4}$ variants of the phase field (i.e., regions with $\eta_i=1$) are represented in blue, green, yellow,
and red, respectively, and the concentration field is represented by solid, dashed, and dotted contour lines at
values $0.35$, $0.5$, and $0.65$, respectively. In all cases, the microstructure evolves as expected in an Ostwald ripening
simulation. Early in the simulation, several particles of each variant exist and interact with one another. 
Later on, some particles disappear and others of the same variant merge. 
Finally, after a sufficiently long time, the microstructure reaches steady state with only one $\eta_{1}$ particle
remaining. The phase evolution results obtained with the different methods are nearly indistinguishable from each
other -- thus illustrating that the negligible error levels assessed at $t=100$ in later Section~\ref{sec:resu:accu} can
be generalized to the entire duration of the simulation.

\newpage

\begin{figure}[H]
  \begin{center}
    \includegraphics[width=\textwidth]{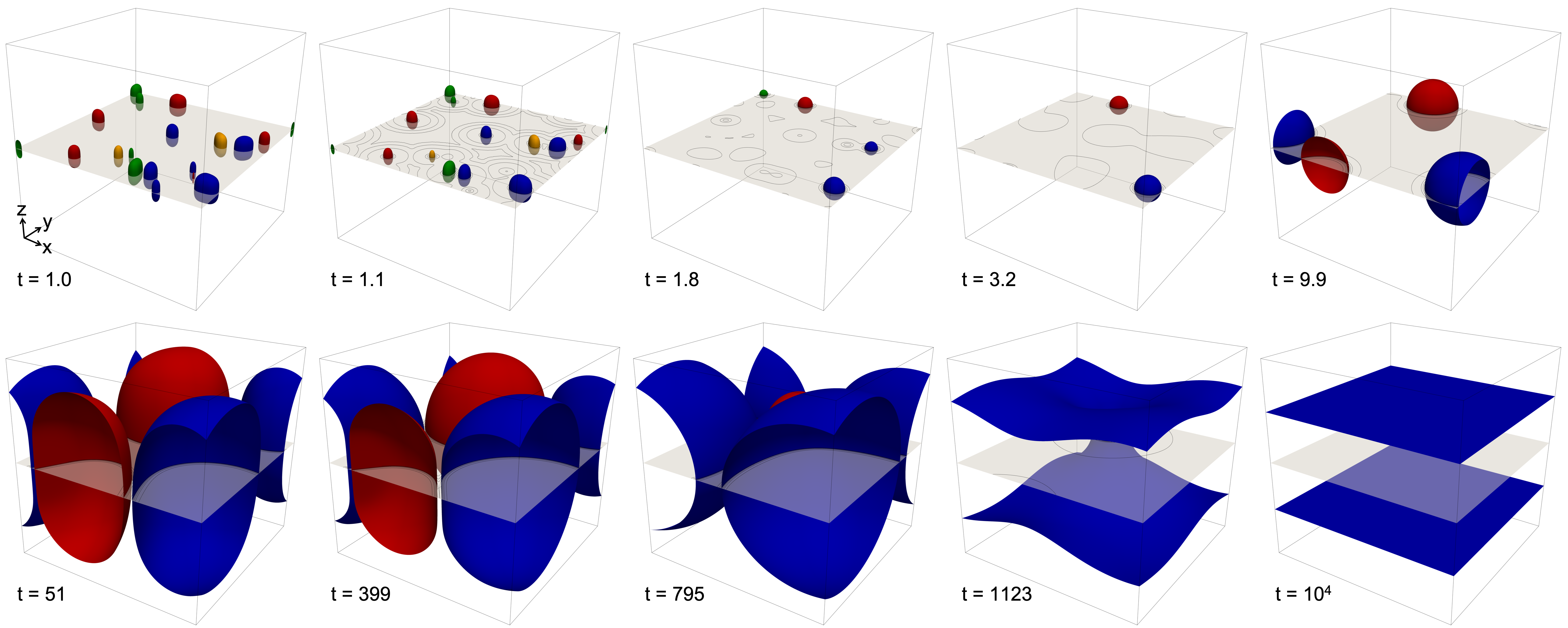}
    \caption{Evolution of $(\eta_i=0)$ interfaces (color) and concentration (isovalue lines along the $(z=l_z/2)$ plane) fields, for a 3D
      simulation with $n=256^3$ obtained with FS method.}
    \label{fig:panel_microstructure_3d}
  \end{center}
\end{figure}

Figure \ref{fig:panel_microstructure_3d} shows the evolution of the 3D microstructure simulated on a $256^3$ grid using the FS method.
As in Fig.~\ref{fig:panel_microstructure_2d}, the $(\eta_i=0.5)$ interfaces for $i=1$, 2, 3, and 4 are represented in blue, green,
yellow, and red, respectively. Isovalues of the concentration field along the central $(z=l_z/2)$ plane are represented from $c=0.3$
to $c=0.8$ with steps of 0.05. In all cases, the microstructure evolves as expected in an Ostwald ripening simulation. Several
small particles of each variant exist at early times and one big (here, planar) particle of $\eta_{1}$ remains when the microstructure
reaches steady state. The microstructure evolves faster in the 3D cases because the domain size and particles are relatively smaller
than in 2D cases.

\subsection{Computational performance}
\label{sec:resu:perf}

\begin{figure}[H]
  \begin{center}
    \includegraphics[scale=0.50]{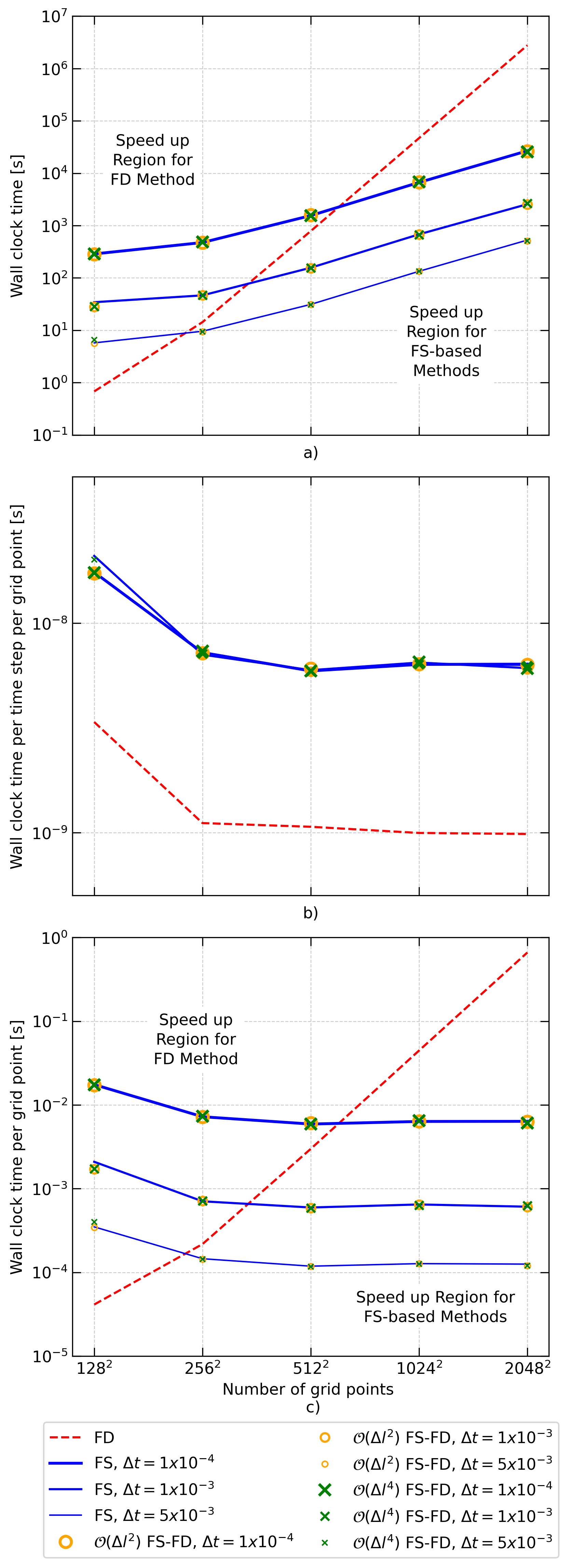}
    \caption{Computational performance for 2D cases. a) wall clock time, b) wall clock time per time step per grid
      point, and c) wall clock time per grid point, all of them for different numbers of grid points.}
    \label{fig:performance_2d}
  \end{center}
\end{figure}

The wall clock time required to complete the 2D simulations from time $1$ to $100$ (dimensionless) is presented in
Figure \ref{fig:performance_2d}.a. For each studied case, 6 simulations are run with the same parameters and the wall clock
time is determined by averaging the results because a variation in the wall clock time consumed by Scikit-CUDA is observed.
Different line thicknesses or symbol sizes represent different $\Delta t$
(thicker/bigger: lower $\Delta t$). Solid blue lines show the FS algorithm, symbols show the FS-FD algorithm
with $\mathcal{O}(\Delta l^{2})$ (yellow $\circ$) or $\mathcal{O}(\Delta l^{4})$ (green $\times$), and dashed
red lines show the corresponding FD algorithm. As expected, the wall clock time decreases with $\Delta t$ due
to the fact that fewer steps are necessary to reach $t=100$. For a given $\Delta t$, the performance of all FS
methods is similar because $\|\boldsymbol{\xi}\|^2$ is computed only once at the initialization stage.

The wall clock time to compute one time step per grid point is shown in Figure~\ref{fig:performance_2d}.b for 2D simulations.
This metric is independent of the FS method and the chosen time step. 
For all methods, it is significantly higher for $n=128^{2}$, which is attributed to parallelization inefficiency when the grid size is small. 
Indeed, a $128^2$ domain corresponds to less
than 3.8 grid points per CUDA core, which is not optimal in the absence of any load balancing supervision. The required
solution time per grid point per time step using the FS methods is between 5.2 and 6.5 times higher than using the FD
method. This is due to the required computation, at every step, of $\mathcal{F}$ of the partial derivatives of $f_{chem}$ at
time $t$, and $\mathcal{F}^{-1}$ of $\widehat{c}$ and $\widehat{\eta_{i}}$ at time $t+\Delta t$, which are not required in FD.
Hence, FS methods perform best when $\Delta t$ is sufficiently larger than with the FD method, namely by a factor
of 6.5 or more, in order to compensate for the extra time needed to calculate $\mathcal{F}$ and $\mathcal{F}^{-1}$. This
compensation is observed at $n \geq 256^{2}$ for $\Delta t= 5 \times 10^{-3}$ (i.e., $\Delta t_{\rm FS}/\Delta t_{\rm FD}=9.8$),
at $n \geq 512^{2}$ for $\Delta t= 10^{-3}$ (i.e., $\Delta t_{\rm FS}/\Delta t_{\rm FD}=28.1$), and at $n \geq 1024^{2}$ for
$\Delta t= 10^{-4}$ (i.e., $\Delta t_{\rm FS}/\Delta t_{\rm FD}=44.9$), as seen in Figures \ref{fig:performance_2d}.a
and \ref{fig:performance_2d}.c. When $n=128^{2}$, the FD method has a better performance in comparison with the FS ones,
but its accuracy is not as good, as discussed in Section~\ref{sec:resu:accu}.

\begin{figure}[H]
  \begin{center}
    \includegraphics[scale=0.50]{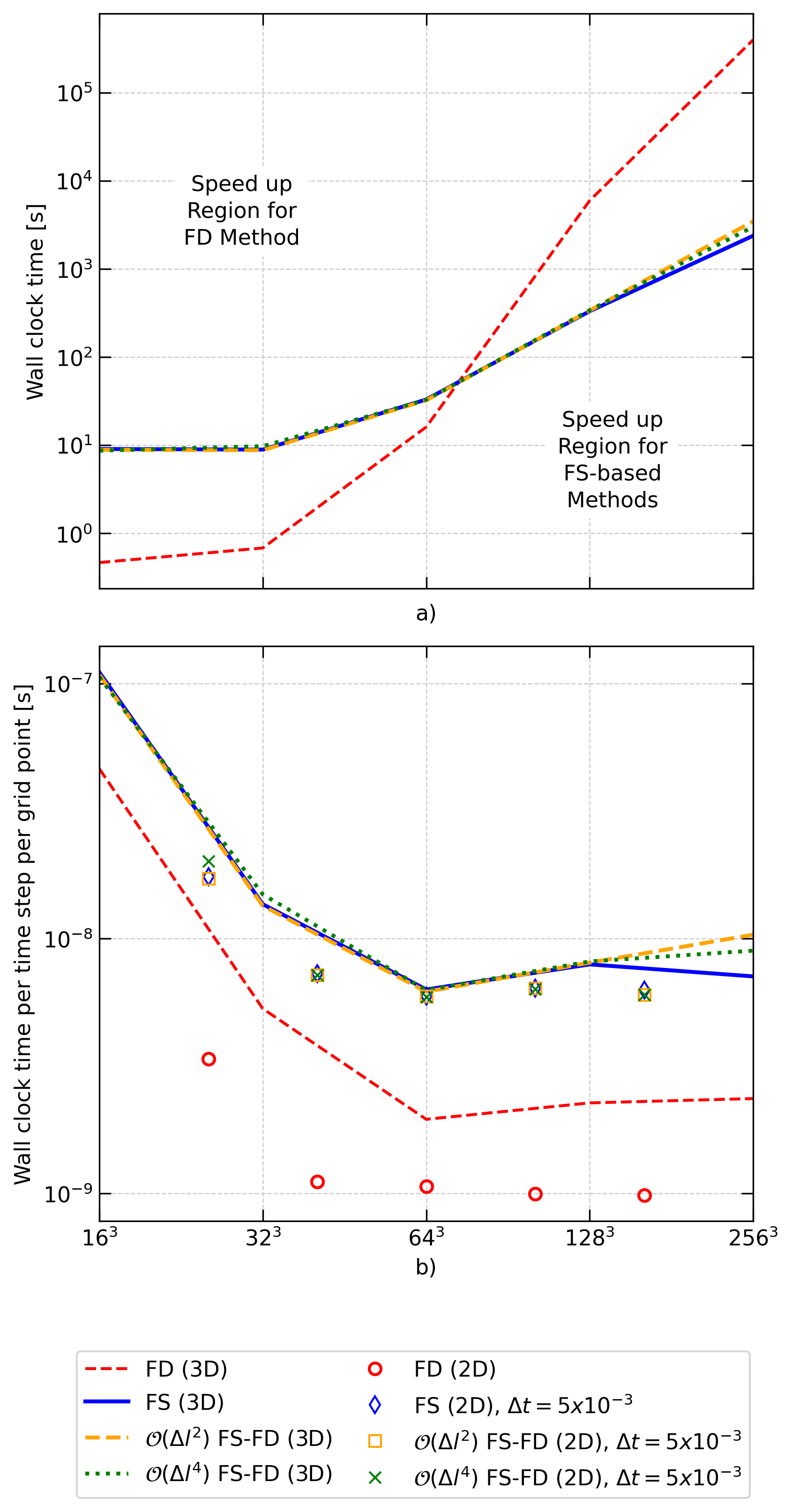}
    \caption{Computational performance for 3D cases for different numbers of grid points (see Table~\ref{tab:cases_3d}):
      a) wall clock time and b) wall clock time per time step per grid point.}
    \label{fig:performance_3d}
  \end{center}
\end{figure}

The wall clock time required to complete the 3D simulations from time $1$ to $100$ (dimensionless) is presented
in Figure \ref{fig:performance_3d}.a. Solid blue lines show the FS algorithm, dashed yellow lines show the
FS-FD algorithm with $\mathcal{O}(\Delta l^{2})$, dashed green lines show the FS-FD algorithm with
$\mathcal{O}(\Delta l^{4})$, and dashed red lines show the corresponding FD algorithm. For each case with
$n=128{3}$ and $n=256^{3}$, computed with FS-based methods, the wall clock time is determined by averaging
the results obtained from 6 simulations because the wall clock time consumed by Scikit-CUDA presents some variations.
The same behavior as in 2D cases is observed and the performance is in favor of FS methods when $\Delta t$ is sufficiently
larger than with the FD method, by a factor of 4.3 or more, in order to compensate for the extra time needed
to calculate $\mathcal{F}$ and $\mathcal{F}^{-1}$. This compensation is observed at $n \geq 128^{3}$
($\Delta t_{\rm FS}/\Delta t_{\rm FD}=62.9$). When $n < 128^{3}$, the FD method has a better performance in
comparison with the FS ones, but its accuracy is not as good. The wall clock time to compute one step per each
grid point is shown in Figure \ref{fig:performance_3d}.b. This metric is independent of the FS method and
the chosen time step. In all cases, it increases significantly for $n=16^{3}$ and $n=32^{3}$, which is
related to parallelization inefficiency when the grid size is small. The required solution time per time
step per grid point using the FS methods is between 2.3 and 4.3 times higher than the using the FD method.
This ratio is lower than in 2D cases because FS methods compute $\|\boldsymbol{\xi}\|^{2}$ at the
begining of the simulation and in this way the number of operations per grid point is the same for 2D and 3D,
but it is not the case for FD method, which uses 5-point stencil in 2D and 7-point stencil in 3D, increasing
the computational cost per grid point, as illustrated in Figure \ref{fig:performance_3d}.b with 2D cases shown as symbols.

Regarding GPU memory (RAM) consumption, the FS and FD 2D simulations with $n=2048^{2}$ require 1.93~GB
and 0.46~GB, respectively, and the FS and FD 3D simulations with $n=256^{3}$ require 7.14~GB and 1.52~GB,
respectively. This difference arises from the fact that 12 complex arrays are needed for the FS scheme
compared to the 6 real ones for FD. The increase in performance (and accuracy, as shown in Section~\ref{sec:resu:accu})
provided by FS methods thus comes at the cost of a higher memory consumption – but even the 7~GB required by the largest $256^3$
cases with 5 tracked fields easily fits within almost modern GPU.

\subsection{Accuracy of the results}
\label{sec:resu:accu}
The accuracy of results obtained for different values of $\Delta t$ and $n$ is quantified, in comparison to a reference
solution by measuring the L$_2$ normalized global error of a field $\phi$ computed at $t=100$ as:

\begin{equation}
  E_{\phi}=\dfrac{\int_{V} (\phi_{0}-\phi)^{2}\,dV}{\int_{V} \phi_{0}^{2}\,dV}
  \label{eq:error_field}
\end{equation}
where $\phi_{0}$ is the reference solution field and $\phi$ is the assessed solution field for a given test condition.

In the absence of a reference (e.g., analytical) solution, the reference solution for 2D simulations is chosen as the FS implementation
with the lowest $\Delta t=10^{-4}$ and the highest $n=2048^{2}$. For the phase fields, the solution of $\eta_{i}$ ($i=1 \text{ to } 4$) are
added together into a field $\eta$. We only compare accuracy for the 2D calculations, as we consider that the finest $256^3$ grid is
not fine enough to be considered a reference, and we observed trends observed in 2D to be generalizable to 3D.

\begin{figure}[H]
  \begin{center}
    \includegraphics[scale=0.5]{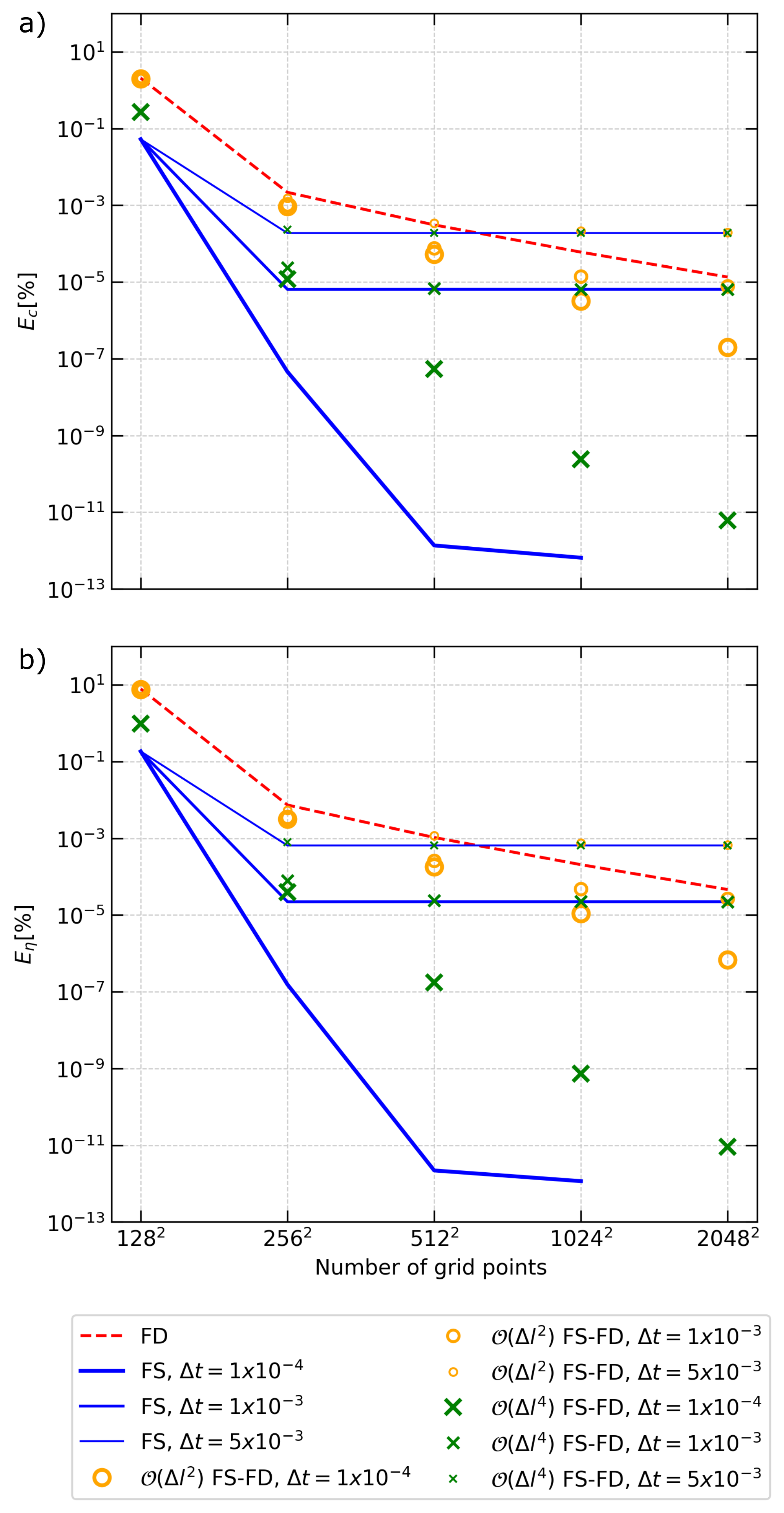}
    \caption{Error for 2D cases of a) concentration field and b) phase field for different numbers of grid point.}
    \label{fig:error_fields}
  \end{center}
\end{figure}

Figure \ref{fig:error_fields} shows the errors, in percentage, for (a) concentration and (b) phase fields with different
numbers of grid points and time step size for 2D cases. As expected, the errors decrease with increasing number of grid
points and decreasing $\Delta t$. 
For the coarsest $n=128^{2}$ grid, good accuracy (i.e., an error lower than 1\%) is achieved, except for FD and
$\mathcal{O}(\Delta l^{2})$ FS-FD methods, giving errors of $E_{c}[\%]=2.03$ and $E_{\eta}[\%]=7.74$.
FS and $\mathcal{O}(\Delta l^{4})$ FS-FD methods provide more accurate results than FD or $\mathcal{O}(\Delta l^{2})$
FS-FD because they lead to more accurate spatial derivatives even on coarser grids. Comparing two cases with equivalent
errors, namely $E_{c}[\%]\approx 10 ^{-6}$ and $E_{\eta}[\%]\approx 2 \times 10 ^{-5}$, the wall clock time ratio between FS
($n=256^{2}$ and $\Delta t=10 ^{-3}$) and FD ($n=2048^{2}$ and $\Delta t=1.490 \times 10^{-7}$) is over four orders of
magnitude ($6\times10^4$), at 46 seconds and 32 days, respectively. Using the $\mathcal{O}(\Delta l^{4})$ FS-FD algorithm,
a lower time step $\Delta t=10 ^{-4}$ is required to achieve a similar accuracy. Since the required time step is one order
of magnitude lower than with FS, the resulting speed-up compared to FD at a similar error is one order of magnitude lower
($\mathcal{O}(\Delta l^{4})$ FS-FD is still faster than FD by over three orders of magnitude). 

In summary, for simulations requiring a sufficient level of accuracy, the FS method, and to a less extent the $\mathcal{O}(\Delta l^{4})$ FS-FD method,
are significantly more efficient than the classical FD algorithm, due to less stringent requirements on both spatial and temporal
discretization.

\section{Conclusions}
\label{sc}
Here, we compared different numerical resolution methods for a benchmark Ostwald-ripening phase-field benchmark simulation.
Our spectral solver makes use of a semi-implicit Fourier spectral-based numerical method, implemented in
Python language, and parallelized on a single GPU. By comparison with first-order forward Euler finite difference
for time discretization and second-order centered finite difference for space discretization, also parallelized on
a single GPU, we conclude that:

\begin{itemize}
\item{Fourier spectral-based methods significantly outperform finite differences
  when a large number of spatial grid points is required. This advantage is due to
  the much larger stable time step afforded by semi-implicit FS methods against explicit FD, in spite of the 
  additional operations (transformation and anti-transformation of complex/scalar variables) per time step
  required by the FS solver. For our implementation, the performance is in favor of FS methods as long as $\Delta t$ is
  larger than the maximum stable time step for the FD method by a factor of 6.5 in 2D or 4.3 in 3D.} 
\item{The time step stability does not strongly depend on the grid size in the semi-implicit
  FS-based methods, which represents a significant advantage when a fine 
  spatial discretization is required.}
\item{Under the same temporal and spatial discretization conditions, all of the Fourier
  spectral-based methods (namely FS or FS-FD) have the same computational performance, but the highest
  accuracy is obtained with the Fourier spectral scheme.}
\item{For a similar level of accuracy (i.e., of error) for both phase and concentration fields,
  the computational performance of the FS and FS-FD $\mathcal{O}(\Delta l^4)$ methods exceeds that of explicit FD by more than
  four orders of magnitude.}
\item{The Python programming allowed an easy implementation of the model
  and exploitation of the benefit of GPU device through the CUDA kernel implementation}.
\item{For the consider benchmark, Fourier spectral-based methods required 4.7 times more memory (RAM) than the
  explicit FD, which may limit the applicability of the former for 3D domains with fine grids using single-GPU hardware.}
\end{itemize}

\section*{Acknowledgements}
This research was supported by the Science Foundation Ireland (SFI) under grant number 16/RC/3872.
D.T. acknowledges the financial support from the Spanish Ministry of Science through the Ramon y Cajal grant RYC2019-028233-I.
A.B. acknowledges the support of ECHV and financial support from HORIZON-TMA-MSCA-PF-EF 2021 (grant agreement 101063099).

\section*{Data Availability}
Data will be made available on request. 

\bibliographystyle{ieeetr}
\bibliography{ref}

\end{document}